\documentclass[twocolumn]{aastex62}

\usepackage[utf8]{inputenc}
\usepackage{graphicx,psfrag}
\usepackage{mathrsfs}
\usepackage{amsmath,amsfonts,amssymb}
\usepackage{multirow} 
\usepackage{diagbox}
\usepackage{comment}
\usepackage{xcolor}
\usepackage{enumerate}
\usepackage{booktabs}
\usepackage{lineno}
\usepackage{lipsum}
\usepackage{mathtools}
\usepackage{color,soul}
\usepackage[normalem]{ulem}

\usepackage{hyperref}
\hypersetup{
    colorlinks = true,
    linkcolor = {blue},
    citecolor = {blue},
    urlcolor = {blue},
    linkbordercolor = {white},
    }
   
\newcommand{\Msun}{$M_\odot$ }

\begin{document}

\title{Towards accelerated nuclear-physics parameter estimation from binary neutron star mergers:\\ Emulators for the Tolman-Oppenheimer-Volkoff equations}

\author[0000-0002-7775-5423]{Brendan~T.~Reed}
\affiliation{Theoretical Division, Los Alamos National Laboratory, Los Alamos, NM 87545, USA}

\author[0000-0003-0427-3893]{Rahul~Somasundaram}
\affiliation{Theoretical Division, Los Alamos National Laboratory, Los Alamos, NM 87545, USA}
\affiliation{Department of Physics, Syracuse University, Syracuse, NY 13244, USA}

\author[0000-0002-3316-5149]{Soumi~De}
\affiliation{Theoretical Division, Los Alamos National Laboratory, Los Alamos, NM 87545, USA}

\author[0000-0002-1198-7774]{Cassandra~L.~Armstrong}
\affiliation{Intelligence and Space Research Division, Los Alamos National Laboratory, Los Alamos, NM 87545, USA}

\author[0000-0002-8145-0745]{Pablo~Giuliani}
\affiliation{Facility for Rare Isotope Beams, Michigan State University, East Lansing, Michigan 48824, USA}

\author[0000-0002-0355-5998]{Collin~Capano}
\affiliation{Department of Physics, Syracuse University, Syracuse, NY 13244, USA}
\affiliation{Physics Department, University of Massachusetts Dartmouth, North Dartmouth, MA 02747, USA}

\author[0000-0002-9180-5765]{Duncan~A.~Brown}
\affiliation{Department of Physics, Syracuse University, Syracuse, NY 13244, USA}

\author[0000-0003-2656-6355]{Ingo~Tews}
\affiliation{Theoretical Division, Los Alamos National Laboratory, Los Alamos, NM 87545, USA}

\date{\today}

\reportnum{LA-UR-24-25009}

\begin{abstract}
Gravitational-wave observations of binary neutron-star (BNS) mergers have the potential to revolutionize our understanding of the nuclear equation of state (EOS) and the fundamental interactions that determine its properties. 
However, Bayesian parameter estimation frameworks do not typically sample over microscopic nuclear-physics parameters that determine the EOS.
One of the major hurdles in doing so is the computational cost involved in solving the neutron-star structure equations, known as the Tolman-Oppenheimer-Volkoff (TOV) equations. 
In this paper, we explore approaches to emulating solutions for the TOV equations: Multilayer Perceptrons (MLP), Gaussian Processes (GP), and a data-driven variant of the reduced basis method (RBM). 
We implement these emulators for three different parameterizations of the nuclear EOS, each with a different degree of complexity represented by the number of model parameters. 
We find that our MLP-based emulators are generally more accurate than the other two algorithms whereas the RBM results in the largest speedup with respect to the full, high-fidelity TOV solver.
We employ these emulators for a simple parameter inference using a potentially loud BNS observation, and show that the posteriors predicted by our emulators are in excellent agreement with those obtained from the full TOV solver. \\
\end{abstract}

\section{Introduction}
\label{sec:intro}

Multi-messenger observations of neutron stars (NS), such as the detection of the binary neutron-star (BNS) merger GW170817~\citep{LIGOScientific:2017ync,LIGOScientific:2017vwq}, NS mass observations~\citep{Demorest:2010bx,Antoniadis:2013pzd,NANOGrav:2019jur,Fonseca:2021wxt}, and observations of NSs by NASA's Neutron Star Interior Composition Explorer~\citep{Riley:2019yda,Miller:2019cac,Riley:2021pdl,Miller:2021qha}, have given us fascinating new insights into the equation of state (EOS) of the densest matter in the universe~\citep{Annala:2017llu,Capano:2019eae,Raaijmakers:2019qny,Essick:2019ldf,Malik:2018zcf,Al-Mamun:2020vzu,Tews:2018iwm,Dietrich:2020efo,Guven:2020dok,Landry:2020vaw,Huth:2021bsp,Koehn:2024set,Breschi:2024qlc}. 
Nevertheless, several questions regarding the properties of dense matter~\citep{Lattimer:2004pg}, such as the possibility of a QCD phase transition in the inner core of NSs~\citep{Annala:2019puf,Somasundaram:2021clp}, remain unanswered. 
The advent of the next-generation of gravitational-wave (GW) detectors, such as Cosmic Explorer~\citep{Reitze:2019iox,Evans:2021gyd} and the Einstein Telescope~\citep{Punturo:2010zz,Branchesi:2023mws}, is expected to dramatically improve our constraints on the EOS, thereby providing us with key information to answer these open questions in nuclear astrophysics~\citep{Finstad:2022oni,Bandopadhyay:2024zrr,Rose:2023uui}.

Prospects for measuring the EOS with gravitational waves have been studied in the literature, in the context of both present~\citep{LIGOScientific:2018cki,Capano:2019eae,De:2018uhw,Annala:2017llu,Malik:2018zcf,Essick:2019ldf,Kunert:2021hgm} and next-generation~\citep{Bandopadhyay:2024zrr,Rose:2023uui,Walker:2024loo,Prakash:2023afe,Landry:2022rxu,Iacovelli:2023nbv} GW detectors. 
However, these studies typically do not sample over nuclear physics parameters, something that is required to comprehensively probe the underlying nuclear physics from astrophysical observations.  
Instead, hierarchical inference~\citep{Walker:2024loo,HernandezVivanco:2019vvk,Essick:2019ldf,Landry:2018prl} is often used to first obtain the marginalized likelihood on the NS component masses and tidal deformabilities, which is then integrated over different EOS curves to ultimately yield constraints on the EOS. 
Alternatively, some studies~\citep{Capano:2019eae,Dietrich:2020efo,Pang:2022rzc} pre-tabulate large numbers of EOSs and sample over an EOS index during the parameter inference run. 
Both approaches have shortcomings. 
The former does not incorporate density-dependent theoretical uncertainties in the EOS models when the parameter inference is performed.
The latter approach does not suffer from this drawback, but might be unsatisfactory for calculating posterior distributions over EOS parameters.
This is because this approach maps one sampling parameter -- the EOS index -- to several nuclear physics parameters. 

To overcome these problems, it is important to sample over the nuclear-physics parameters simultaneously with the other binary parameters. 
This implies an ``on-the-fly'' generation of the EOS and its resulting global NS properties.
Hence, every likelihood evaluation in such a Bayesian inference run requires solving the Tolman–Oppenheimer–Volkoff~\citep{Oppenheimer:1939ne} (TOV) equations.
The TOV equations are the NS structure equations that relate the EOS -- and hence, the nuclear-physics parameters -- to astrophysical observables such as NS radii $R$ and tidal deformabilities $\Lambda$~\citep{Hinderer:2009ca} as function of masses $M$. 
A typical parameter inference run for a BNS event requires $\approx 10^7$ likelihood evaluations~\citep{Veitch:2014wba}. 
Therefore, given that a single solution to the TOV equations takes up to a few seconds (when also solving for $\Lambda$), employing the full high-fidelity TOV solver is prohibitively expensive. 
Furthermore, this problem will be exacerbated in the era of third-generation GW detectors, which are expected to detect $\mathcal{O}(1000)$ high-precision BNS events per year~\citep{Evans:2021gyd}.
This large number of observations will provide unparalleled information to constrain nuclear-physics parameters if these can be sampled over directly.
Finally, rapid parameter estimation from GW signals will aid efficient electromagnetic follow-up~\citep{Margalit:2019dpi}.

So far, EOS parameters were directly sampled over only for a single event, GW170817~\citep{LIGOScientific:2018cki}, using a relatively simple spectral decomposition model for the EOS~\citep{Lindblom:2010bb}.
Due to the above-mentioned computational expense, sampling over EOS parameters ``on-the-fly'' has not been attempted for more complicated EOS models whose parameters are motivated by nuclear theory and experiment.
Here, we present a novel approach to this problem by employing emulators~\citep{Canizares:2014fya,Smith:2016qas,Bonilla:2022rph}, i.e. algorithms that mimic solutions to the TOV equations but at a fraction of their computational cost.
Recently, such emulators have been applied to NS structure equations~\citep{Tiwari:2024jui,Liodis:2023adg}.
However, these works have been limited to emulating NS radii, and have focused on emulators based on neural networks. 
Instead, we develop three different approaches to TOV emulators: 
a neural network-based emulation using Multilayer Perceptrons (MLP), 
Gaussian Processes (GP), 
and a data-driven variant of the reduced basis method (RBM).
While our algorithms can be used to emulate any global NS property, in this paper, we focus on emulating the NS tidal deformability $\Lambda$~\citep{Hinderer:2009ca} because it is directly relevant for GW observations of binary NS mergers.   
We explore the performance of our emulators for three different EOS models with varying degree of complexity: a 1-parameter, 5-parameter, and 10-parameter model. 
Both the 5-parameter and 10-parameter models are general enough to account for constraints from nuclear theory and experiment at low densities, while being agnostic enough to allow for phase transitions at higher densities. 
We find that for all EOS sets, our MLP emulator is the most accurate of the three algorithms and the RBM is the fastest.
Using a mock BNS event, we demonstrate the power of our emulators by comparing the probability density functions (PDFs) on nuclear-physics parameters predicted by our emulators with those obtained by using the full TOV solver. 

The paper is organized as follows. 
In Sec.~\ref{sec:EOS}, we detail the construction of our three parametric EOS models and of both training and validation sets for the emulators.
In Sec.~\ref{sec:methods}, we lay out our implementation of the three emulators: MLP (Sec.~\ref{sec:MLP}), GP (Sec.~\ref{sec:GP}), and RBM (Sec.~\ref{sec:RBM}). 
We also evaluate the performance of our emulators by validating them against a large set of pre-generated EOS samples. 
Section~\ref{sec:CAT} compares the three emulators, considering both their accuracy and speed.
Finally, in Sec.~\ref{sec:loud_event}, we study the ability of our emulators to obtain accurate PDFs from a typical BNS event that could be detected by the next-generation of GW detectors.
Our conclusions are presented in Sec.~\ref{sec:conclusion}.

\section{EOS models and data generation}
\label{sec:EOS}

\begin{figure*}
    \centering
    \includegraphics[width=0.99\columnwidth]{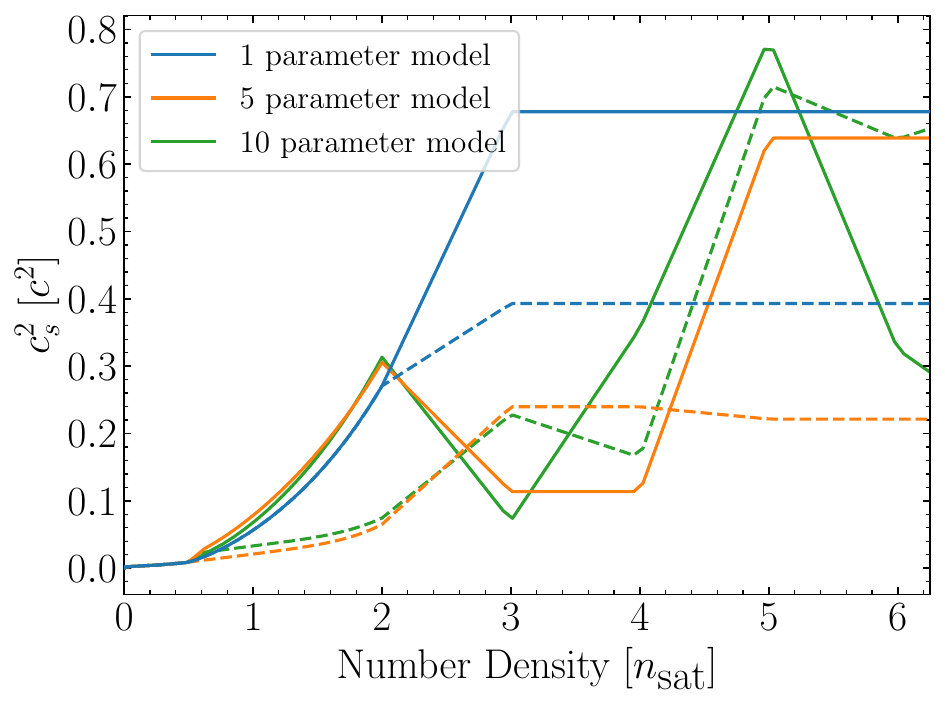}
    \includegraphics[width=0.99\columnwidth]{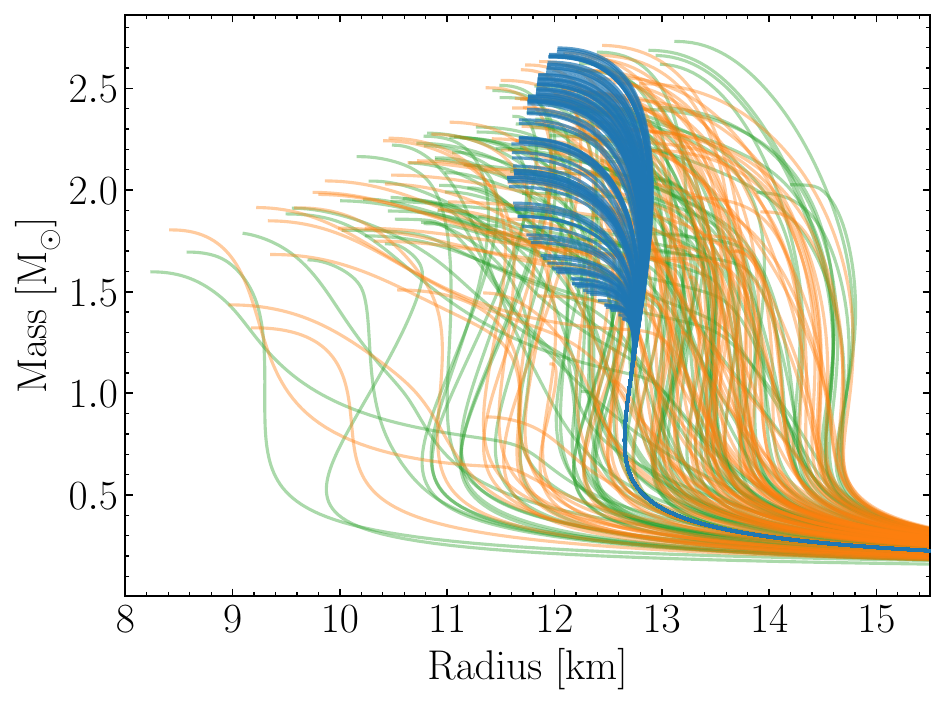}
    \caption{(Left) The density-dependent sound-speed profiles for two samples for each of our three EOS models in units of $c^2$, where $c$ is the speed of light. 
    (Right) The resulting mass-radius curves for 100 samples of each EOS model.}
    \label{fig:EOS_samples}
\end{figure*}

Models for EOS that are employed in GW inference analyses need to be versatile enough to provide sufficiently broad priors on astrophysical NS observables and sufficiently flexible so that they can be connected in a straightforward manner to experimental and theoretical nuclear-physics inputs.
Common examples of such EOS models are the piece-wise polytrope model~\citep{Read:2008iy,Hebeler:2013nza}, the spectral decomposition approach~\citep{Lindblom:2010bb}, Gaussian Processes~\citep{Essick:2019ldf,Landry:2020vaw}, and the speed-of-sound model~\citep{Tews:2018kmu,Greif:2018njt}. 
Here, we employ the EOS modeling approach recently used in~\cite{Koehn:2024set} and \cite{Komoltsev:2023zor}. 

In this approach, the outer core of NSs is assumed to be composed of only nucleonic degrees of freedom.
The EOS in this part is described by the metamodel developed in ~\cite{Margueron:2017eqc} and \cite{Margueron:2017lup}.
The metamodel is an extremely flexible density-functional approach whose model parameters are the so-called nuclear empirical parameters (NEPs)~\citep{Margueron:2017eqc} that govern the behavior of the EOS of pure neutron matter and symmetric nuclear matter.
The NEPs are given by taking the usual expansion of the nuclear matter EOS in powers of the proton-neutron asymmetry,
\begin{eqnarray}
    E(n,\alpha)=\mathcal{E}(n)+\alpha^2 S_2(n)+\mathcal{O}(\alpha^4)
\end{eqnarray}
where $E(n,\alpha)$ is the total energy per particle, $\mathcal{E}(n)$ is the energy of symmetric matter, $S_2(n)$ is the quadratic approximation to the symmetry energy~\citep{Somasundaram:2020chb}, and $\alpha=\frac{n_n-n_p}{n}$ is the proton-neutron asymmetry parameter. 
Furthermore, the symmetry energy $S(n)$ is defined as the difference between the energy of pure neutron matter and symmetric nuclear matter, i.e. $S(n) \equiv E(n,\alpha=1)-\mathcal{E}(n)$.   
The energy of symmetric matter and the symmetry energy are then expanded around nuclear saturation density $n_{\rm sat}$
\begin{eqnarray}
    &&\mathcal{E}(n)=E_\text{sat}+\frac{1}{2}K_\text{sat}x^2+\frac{1}{6}Q_\text{sat}x^3+\frac{1}{24}Z_\text{sat}x^4 + \dots \\
    &&S(n)=E_\text{sym}+L_\text{sym}x+\frac{1}{2}K_\text{sym}x^2+\frac{1}{6}Q_\text{sym}x^3 \nonumber \\ && \quad \quad \quad +\frac{1}{24}Z_\text{sym}x^4 + \dots
\end{eqnarray}
where $x=\frac{n-n_\text{sat}}{3n_\text{sat}}$.
It has been shown that the metamodel is capable of reproducing the EOS of a large number of nucleonic models via a suitable adjustment of the NEPs. 
Constraints from microscopic ab-initio nuclear theory, e.g., from chiral EFT~\citep{Hebeler:2013nza,Tews:2018kmu,Drischler:2017wtt,Keller:2022crb}, and nuclear experiment, such as PREX~\citep{Reed:2021nqk}, can be incorporated in the form of suitable priors on the NEPs~\citep{Koehn:2024set}.

The description in terms of nucleons might break down at higher densities.
Hence, as in \cite{Koehn:2024set}, we employ the speed-of-sound model of \cite{Tews:2018kmu} to describe the EOS above a breakdown density of the metamodel that we chose to be $2 n_\text{sat}$.
For the speed-of-sound model, the parameters are the squared sound speeds at discrete density points. 
Unlike in \cite{Koehn:2024set}, we fix these density points at $3 n_\text{sat}$, $4 n_\text{sat}$, and so on. 
The resulting density-dependant speed of sound can be integrated to obtain the pressure, energy density, and chemical potential.
This can, in turn, be used as input for the TOV equations which yields the NS mass, radius and tidal deformability, see ~\cite{Tews:2018kmu} and \cite{Somasundaram:2021clp} for more details.

We employ our full EOS model in various degrees of complexity to train our emulators. 
For our 1-parameter model, we fix all the NEPs: $E_\text{sat}=-16$~MeV, $n_\text{sat}=0.16$~fm$^{-3}$, $K_\text{sat}=230$~MeV, $Q_\text{sat}=Z_\text{sat}=0$~MeV, $E_\text{sym}=32$~MeV, $L_\text{sym}=50$~MeV, and $K_\text{sym}=Q_\text{sym}=Z_\text{sym}=0$~MeV. 
Hence, the EOS up to $2 n_\text{sat}$ remains unchanged. 
Then, we assume the speed of sound above $3 n_\text{sat}$ to be constant and use its value as the one free parameter in the model; see left panel of Fig.~\ref{fig:EOS_samples}. 
For our 5-parameter model, we vary the NEPs $L_\text{sym}$, $K_\text{sym}$ and $K_\text{sat}$, with the other NEPs fixed at the same values as for the 1-parameter model. 
Additionally, the speed of sound at $3 n_\text{sat}$ and $5 n_\text{sat}$ are free parameters. 
In this model, the speed of sound is chosen to be constant between $3 n_\text{sat}$ and $4 n_\text{sat}$, and above $5 n_\text{sat}$.  
Finally, we construct a 10-parameter model which has the same level of flexibility for the EOS below $2 n_\text{sat}$, but varies the $7$ sound speeds at $3 n_\text{sat}$, $4 n_\text{sat}$, $\dots 9 n_\text{sat}$.
For all models, all grid points are connected using linear segments. 
The left panel of Fig.~\ref{fig:EOS_samples} illustrates the construction of our EOS models by showing the sound speed profile for two samples per model.

To train our emulators, we generate a set of 100,000 samples for each EOS model and a separate set of 100,000 samples per model for validation.
The EOS samples are obtained by varying the model parameters in a uniform range. 
All the (squared) sound-speed parameters are sampled uniformly between $0$ and $1$ in units of the speed of light, whereas we use the following ranges for the NEPs: $L_\text{sym}=[20,150]$~MeV, $K_\text{sym}=[-300,100]$~MeV, and $K_\text{sat}=[200,300]$~MeV.
For each EOS sample, we solve the TOV equations to obtain masses, radii, and tidal deformabilities. 
We show 100 resulting mass-radius curves for each EOS model in the right panel of Fig.~\ref{fig:EOS_samples}. 
Finally, for both the training and validation sets, we discard the samples that do not reach $2$M$_\odot$~\citep{Demorest:2010bx,Antoniadis:2013pzd,NANOGrav:2019jur,Fonseca:2021wxt}, approximately $30\%$ of each set. 

\section{Emulation Strategy}
\label{sec:methods}

Since our goal is to perform Bayesian parameter inference for BNS mergers by explicitly sampling over EOS model parameters, we train our emulators to take as input the EOS model parameters and output the tidal deformabilities $\Lambda$ for a sequence of NS masses. 
Here, we focus on emulating $\Lambda$ but our framework can be easily extended to other astrophysical observables such as the NS radius. 
In the following, we separately discuss the construction of our three different emulators and their performance as characterized by their validation accuracies.
In the next section, we will compare the emulators in detail regarding their accuracy and speed.

\subsection{Multilayer Perceptron}
\label{sec:MLP}

\begin{figure}
    \centering
    \includegraphics[width=0.99\columnwidth]{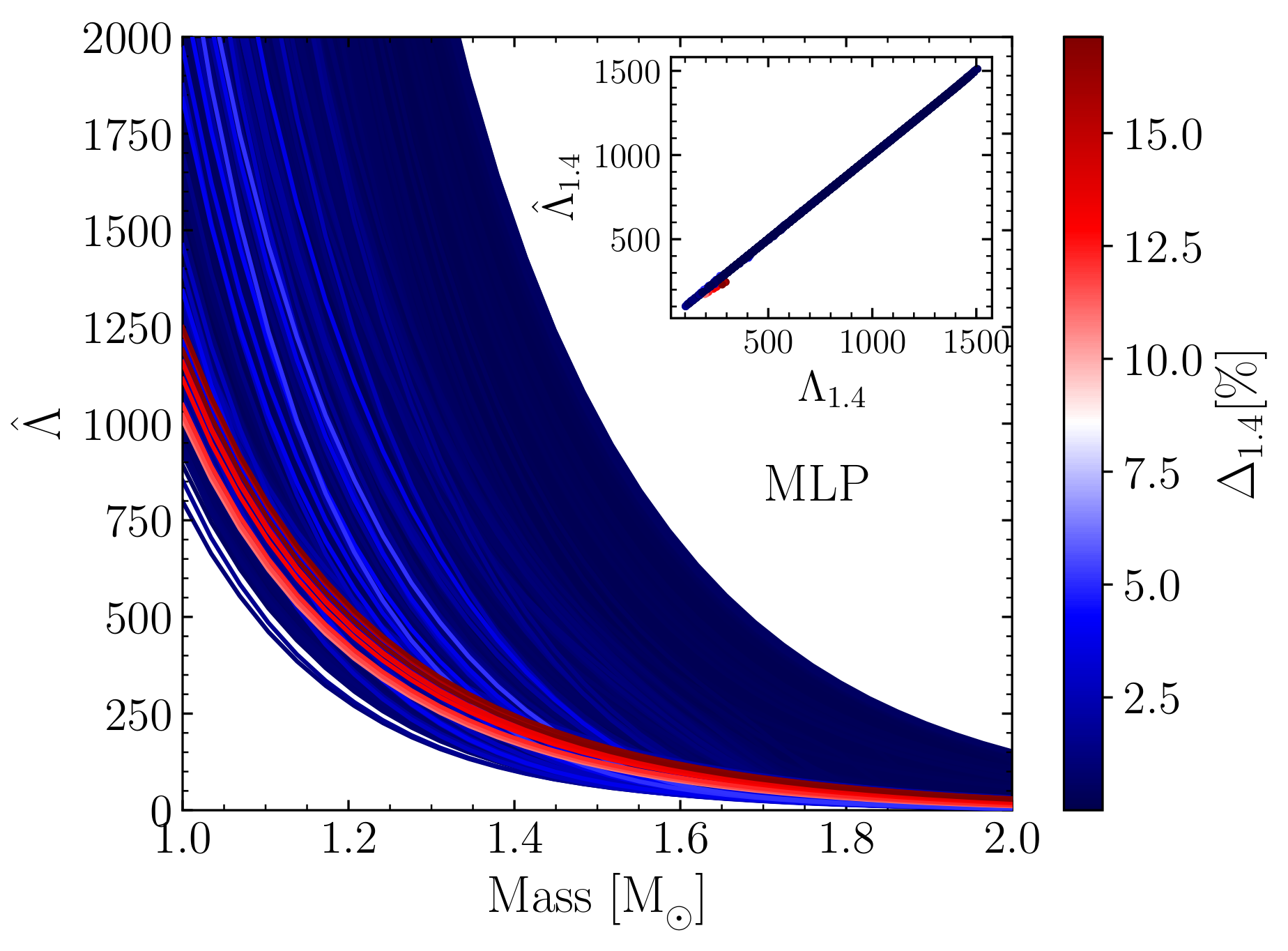}
    \caption{Validation results for the MLP  for the 5-parameter EOS model. 
    We show the predicted tidal deformability $\hat{\Lambda}$ for each sample in the validation set as a function of the NS mass. 
    The color scheme corresponds to the validation percentage uncertainty of a given sample at $1.4$M$_\odot$, $\Delta_{1.4}$.
    The inset depicts the cross-validation result, with the predicted tidal deformability at $1.4$M$_\odot$, $\hat{\Lambda}_{1.4}$, plotted against the corresponding true value for every validation sample. 
    }
    \label{fig:CV_MLP}
\end{figure}

The MLP is considered to be the simplest feedforward neural network, consisting of fully connected neurons~\citep{NN_book}. 
The number of neurons in the input layer is determined by the number of input EOS model parameters, i.e. either $1$, $5$, or $10$ depending on the EOS model. 
For the 1-parameter model, our MLP consists of $15$ neurons in the output layer, with the activation of each neuron in the output layer corresponding to $\log_{10}({\Lambda})$ of a NS with a certain mass. 
Therefore, the MLP effectively predicts the $\Lambda-$~M curve on a fixed mass grid, which we take to be uniformly spaced with 15 points. 
For the 1-parameter model, the upper bound of the grid is $2$M$_\odot$ whereas the lower bound is set to be $1.4$M$_\odot$. 
This is because, below $1.4$M$_\odot$, there is no significant variation in $\Lambda$ with respect to the EOS parameters since the EOS below $2n_\text{sat}$ remains unchanged, see Fig.~\ref{fig:EOS_samples}.
Similarly, for the 5- and 10-parameter models, the output mass grid consists of 30 points uniformly spaced between $1$M$_\odot$ and $2$M$_\odot$.

For the 1-parameter model, we use two hidden layers, each with 64 neurons. 
For the 5 and 10-parameter models, we use 5 hidden layers, each containing 64 neurons. 
These hyperparameters of the MLP architecture were chosen in order to optimize the performance of the MLP.
We found that the MLP's performance accuracy increases as a function of the number of hidden layers but saturates roughly at the chosen values. 
On the other hand, the number of neurons per hidden layer did not significantly impact the behaviour of the MLP and the value $64$ was found to be a good choice. 
The activation function of all hidden neurons in our MLPs is taken to be the rectified linear unit, whereas we use the identity function for the output neurons.

To increase the accuracy of the MLP emulator, we employ the method of Bagging~\citep{Bagging}.
For each EOS model, we construct not one, but 100 MLPs, all with the same architecture and hyperparameters discussed above.
The training of the MLPs is performed using a stochastic gradient-based optimizer, {\sc Adam}~\citep{kingma2017adam}.
Therefore, the training is a non-deterministic process which results in a variance among the predictions of the MLPs. 
The final prediction of the full MLP ensemble is obtained by averaging over all individual MLPs. 
In the following, we will refer to this emulation strategy simply as MLP. 
Bagging $100$ individual MLPs decreases the speed of the full emulator by a factor $100$. 
To investigate the trade-off between speed and accuracy, we also investigate the predictions of a single MLP in this ensemble, which we take to be the one with the lowest residual loss function. 
This emulator is referred to in the following as MLP$^*$. 
All MLPs in the ensemble are trained using all available training samples for all EOS models. 
The MLPs are trained until the loss function remains constant to within $10^{-10}$ for 10 consecutive epochs. 
This typically results in a total of $\approx 100$ epochs for each MLP. 

Fig.~\ref{fig:CV_MLP} shows the validation performance of our MLP emulator. 
In this section, we show results only for the 5-parameter EOS model but results are qualitatively similar for the other sets. 
We find that for the vast majority of validation samples, the percentage uncertainty evaluated at $1.4$M$_\odot$, $\Delta_{1.4}$, is less than $1\%$.
The average of $\Delta_{1.4}$ over all EOS is only $0.04\%$, indicating that the MLP is a faithful emulator of the full TOV solver. 
The most significant outlier EOS has an uncertainty $\Delta_{1.4} \approx 15\%$, but such samples are rare, with only $\approx 10$ out of $\approx 70,000$ samples exhibiting such large uncertainties.
From the cross-validation plot we see that samples with larger $\Delta_{1.4}$ are softer EOSs. 
This is because both the training and validation data are generated by uniform sampling over the EOS model parameters. 
Given the non-linear dependence of the tidal deformability on the EOS parameters, this results in a larger number of stiff EOSs which consequently dominate the training of the MLP.

\subsection{Gaussian Processes}
\label{sec:GP}

Gaussian Process regression~\citep{GP_book,Wang} offers a non-parametric, probabilistic alternative to the MLP.
Similar to the MLP, our GP emulator is designed to accept a set of EOS parameters as input, and predict $\log_{10}(\Lambda)$ for a sequence of NSs on a uniform mass grid. 
As with the the MLP, the mass grid consists of $15$ points uniformly spaced between $1.4$M$_\odot$ and $2$M$_\odot$ for the 1-parameter model, whereas we use $30$ points between $1$M$_\odot$ and $2$M$_\odot$ for the 5 and 10-parameter models. 

The prior GP is specified by the choice of the kernel which we take to be the Matérn kernel~\citep{GP_book}. 
The Matérn kernel is a generalization of the traditional radial basis function kernel and we found the former to generally perform better. 
The hyperparameters of the kernel are optimized by maximizing the log-marginal-likelihood and the posterior GP is obtained using Algorithm~2.1 of ~\cite{GP_book}, as implemented in the python package $\textsc{scikit-learn}$. 
In contrast to MLPs, ordinary GPs, such as those considered in this work, scale poorly with the size of the training set and we were unable to train any GP on more than 5000 training EOSs. 
There are a few approaches in the literature to address this problem, such as sparse GPs~\citep{Sparse_GP}.
Here, instead, we make full use of our large dataset by once again employing the Bagging method: instead of a single GP, we consider an ensemble of GPs, each trained on a distinct set of 5000 samples drawn from the original training set.
For each GP, we use the mean of the posterior for a given set of EOS model parameters and average over the outputs of all GPs in the ensemble in order to obtain the final prediction. 
For the 1-parameter model, we employ an ensemble of 10 GPs while for the 5 and 10-parameter models, we use 20 GPs. 

\begin{figure}
    \centering
    \includegraphics[width=0.99\columnwidth]{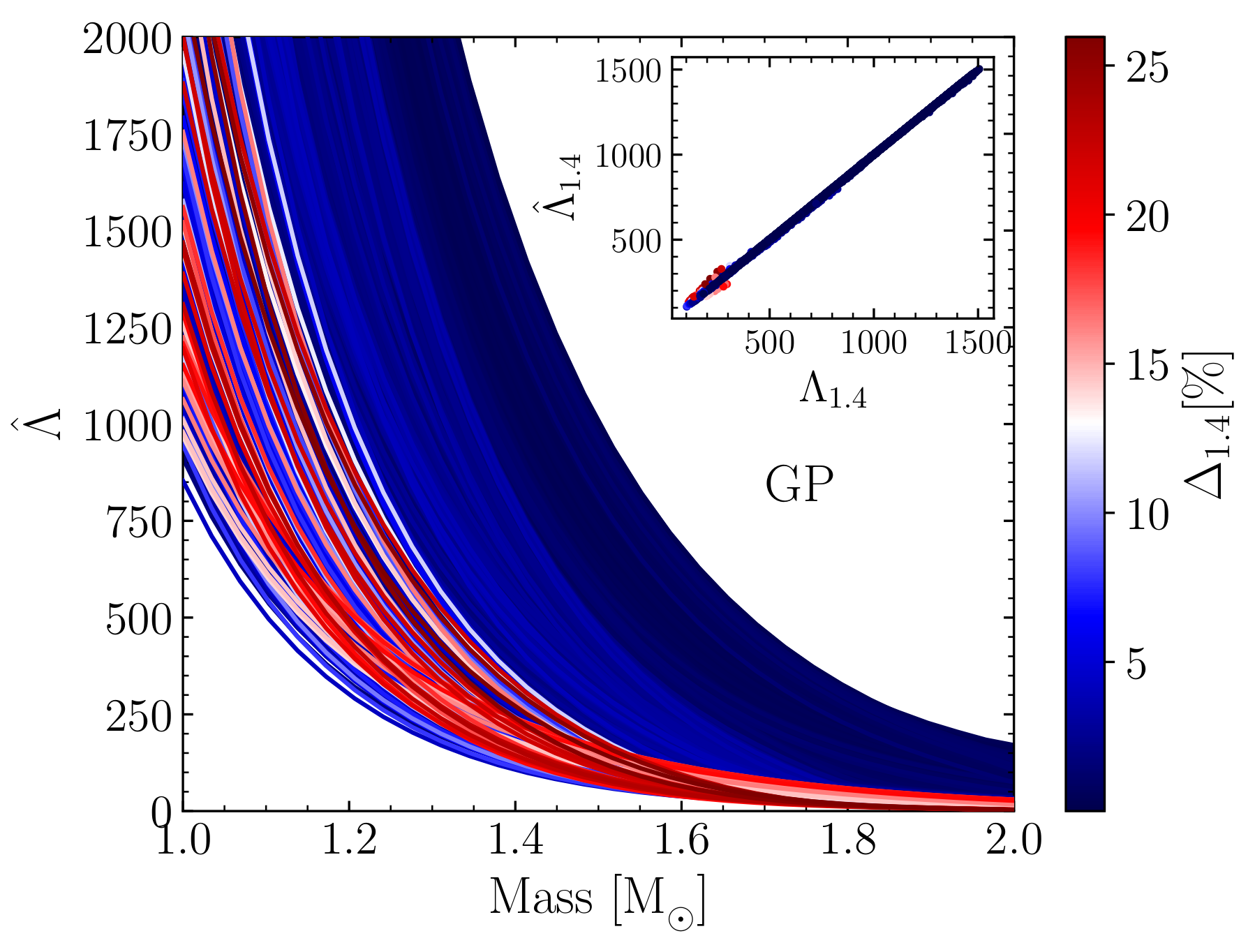}
    \caption{Same as Fig.~\ref{fig:CV_MLP}, but for the GP emulator.}
    \label{fig:CV_GP}
\end{figure}

Fig.~\ref{fig:CV_GP} shows the validation performance of our GP emulator for the 5-parameter EOS model.
We find that, generally, the GP provides a good emulation of the full TOV solver and the average $\Delta_{1.4}$ is $\approx 0.09 \%$.
The most significant outlier has an uncertainty $\Delta_{1.4}\approx 25 \%$ and around 12 samples have such an uncertainty. 
Similar to the results of the MLP emulator, we see that samples with larger uncertainties tend to be softer EOSs.
While a little worse, the GP validation uncertainty is comparable to the MLP emulator and indicates that the GP can be a useful tool for emulating the TOV equations.

\subsection{The Reduced-Basis Method}
\label{sec:RBM}

\begin{figure}
    \centering
    \includegraphics[width=0.99\columnwidth]{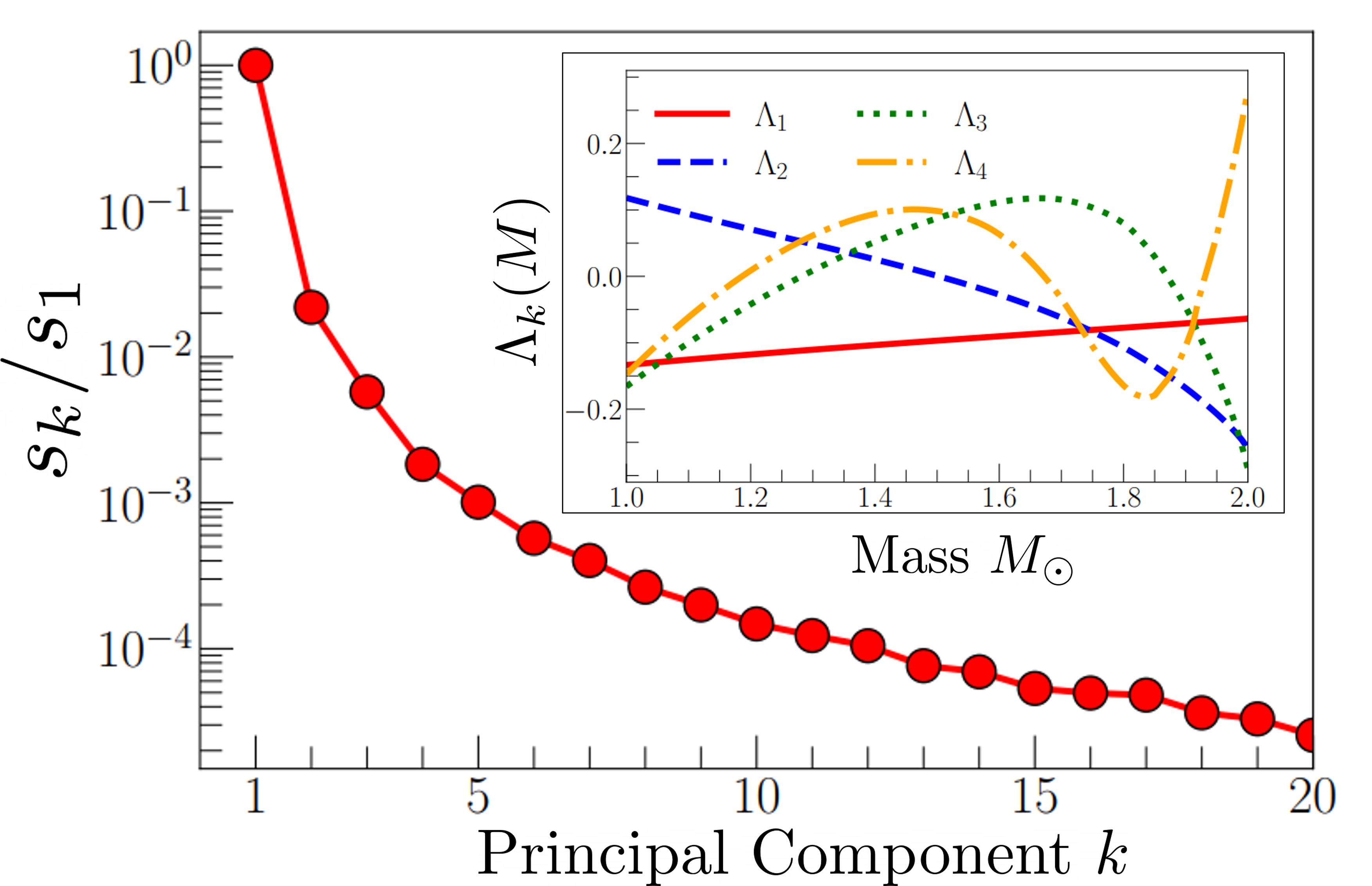}
    \caption{Fast decay of the singular values in the reduced basis expansion for $\log{(\Lambda(M))}$ in~\autoref{eq: reduced coordinates}, which signals that a good approximation can be obtained with just a few components~\citep{Bonilla:2022rph}.
    The inset shows the first four principal components, which represent the first four reduced bases $\Lambda_k(M)$.}
    \label{fig:SVDLambda}
\end{figure}

The reduced basis method (RBM)~\citep{quarteroni2015reduced} is a model-order reduction technique~\citep{quarteroni2014reduced} that aims at emulating a complex system by finding reduced coordinates, usually as coefficients of a small basis expansion, and creating a system of equations for them, usually through a Galerkin projection formalism~\citep{Bonilla:2022rph}.
Even though it has been successfully applied to various problems in nuclear physics~\citep{giuliani2023bayes,rose2024,drischler2023buqeye,Somasundaram:2024zse} and GW astronomy~\citep{field2011reduced,Smith:2016qas}, obtaining the reduced equations from a Galerkin projection is not always a viable option due to the structure of the differential equations involved. 
This is the case for the TOV equations we address here, since they are non-affine in the EOS (pressure, energy density, etc.), and in the controlling EOS parameters.

To overcome this issue, we take a data-driven approach to estimate a set of reduced equations without explicitly calculating the projections~\citep{figueroa2024}, as has been done in various other model-order reduction applications, see for example~\cite{burohman2023data}. 
An alternative approach would rely on the collocation framework~\citep{chen2021eim}, which can still exploit information about the operators involved to construct the reduced equations without the disadvantages of a Galerkin projection scheme. 
This avenue is in preparation and will soon be publicly available~\citep{collocation2024}.

\begin{figure}
    \centering
    \includegraphics[width=0.99\columnwidth]{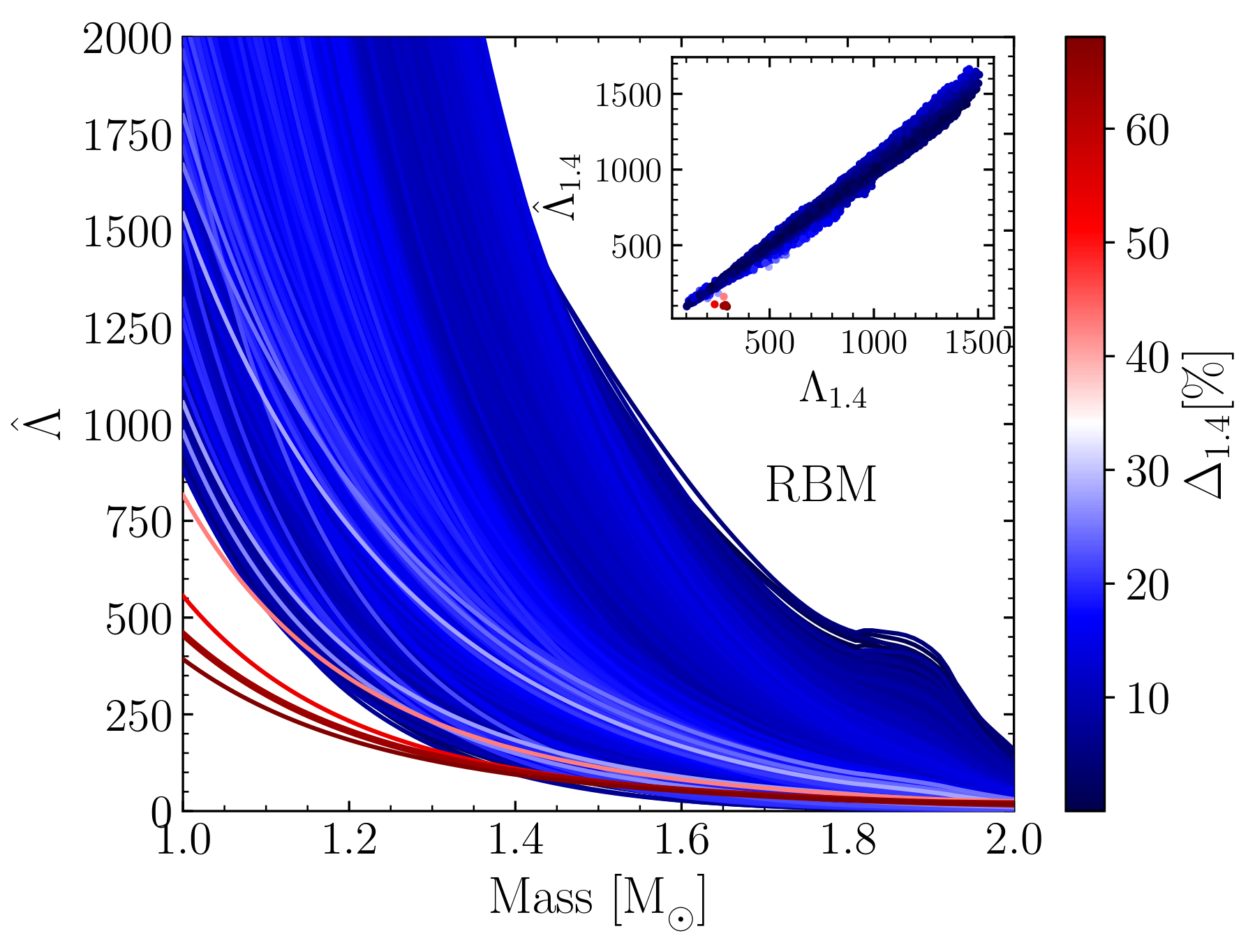}
    \caption{Same as Fig.~\ref{fig:CV_MLP}, but for the RBM emulator.}
    \label{fig:CV_RBM}
\end{figure}

In the following we describe our approach for emulating the tidal deformability $\Lambda(M)$.
The emulation strategy for the radius follows the same procedure but is not discussed here. 
We first construct a reduced basis for the $\log_{10}$ (hereafter just $\log$) of $\Lambda$ as a function of the NS mass and the EOS model parameters $\alpha$:
\begin{equation}\label{eq: reduced coordinates}
    \log{(\Lambda(M;\alpha))}\approx\log{(\hat{\Lambda}(M;\alpha))}=\sum_k^n a_k(\alpha) \Lambda_k(M)\,.
\end{equation}
Here, $\Lambda_k(M)$ represent the reduced bases, and the coefficients $a_k$ the $n$ reduced coordinates that change as we modify the parameters $\alpha$. We use $\log{(\Lambda)}$ in order to provide smoother curves for each reduced basis which will become important for the next step. 
To obtain an optimal basis $\Lambda_k(M)$ we perform a proper orthogonal decomposition~\citep{quarteroni2015reduced}, and construct them as the first $n$ principal components (or singular vectors) of a matrix of snapshots of $\log{(\Lambda(M;\alpha_j))}$ for $j=1,...,\ N$ high fidelity solutions of the TOV equations. 
The snapshot solutions are calculated at a grid of fixed mass $M$ values between 1.0\Msun and 2.0$M_\odot$. Figure~\ref{fig:SVDLambda} shows the decay of the singular values as well as the first four reduced bases for $\log{(\Lambda(M))}$.

\begin{figure*}
    \centering
    \includegraphics[width=0.99\textwidth]{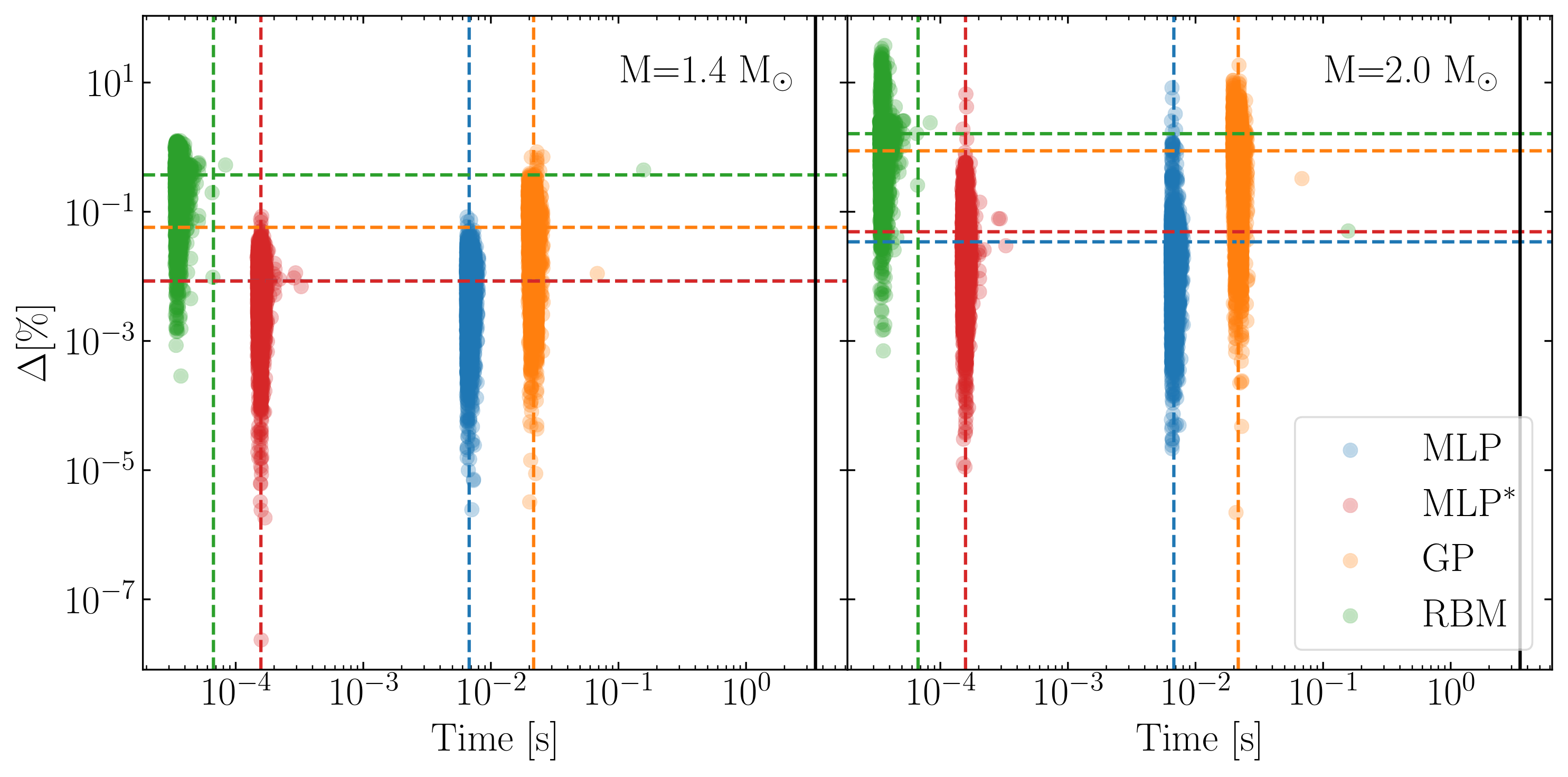}
    \caption{Computational accuracy vs. time plot for the 1-parameter model. This plot helps compare emulators' trade-offs in a similar way to the Pareto front~\citep{brunton2019data}.
    Each dot corresponds to a validation sample and we show a reduced set of only $5000$ validation samples.
    The dashed lines indicate the averages over all $5000$ validation samples.
    The total CPU time required to evaluate the full TOV solver is indicated by the solid black vertical line. 
    Note that in the left panel, the  lines for the MLP and MLP$^*$ overlap.}
    \label{fig:CAT_s1}
\end{figure*}

To obtain the coefficients $\vec a =\{a_k\}$ in the expansion~\eqref{eq: reduced coordinates} as functions of the parameters $\alpha$, we construct implicit equations using an eigensystem of symmetric matrices~\citep{figueroa2024,Cook:2024toj}:
\begin{eqnarray}\label{eq: eigensystem}
    &&H(\alpha)\vec a = \lambda \vec a\,, \\
    &&H(\alpha) = D_0+\sum_i^m f_i(\alpha) H_i\,,
    \label{eq:H-matrix}
\end{eqnarray}
where the diagonal matrix $D_0$, as well as the symmetric matrices $H_i$ have yet undetermined coefficients that are eventually learned from data.
The functions $f_i$ represent \emph{features} of the controlling parameters $\alpha$, such as polynomial expansions of $\alpha$. 
In our case, we choose these features to be the pressure for the targeted EOS evaluated at various densities that are relevant for the NS structure. 
Since solving the system in Eq.~\eqref{eq: eigensystem} gives an eigenvector $\vec a$ that is defined up to a normalization, we further emulate the normalization $N(\alpha)$ by directly using the Parametric Matrix Model (PMM) described in~\cite{Cook:2024toj} with an expansion on analogous features $f_i(\alpha)$\footnote{In our explorations, attempting to encode the normalization in the eigenvalue of the same equation system~\eqref{eq: eigensystem} yielded worse results than training a separate PMM for the normalization alone.}.
The details about the RBM data-driven emulator, including the dimensions for the normalization emulation using the PMM are shown in Table~\ref{tab:lambda-BB}.

\begin{deluxetable}{c||c||c||c||c}[]
\tablehead{\begin{tabular}[c]{@{}c@{}}Parameter\\ Set \end{tabular} & $n$ & $\begin{tabular}[c]{@{}c@{}}PMM\\ dim \end{tabular}$ 
& 
\begin{tabular}[c]{@{}c@{}}$\rho$ (PC)\\  (fm$^{-3}$)  \end{tabular}
& 
\begin{tabular}[c]{@{}c@{}}$\rho$ (Norm)\\  (fm$^{-3}$)  \end{tabular}}
\caption{Parameters of the RBM tidal deformability emulators for each of the three parameters set. The second column denotes the numbers of principal components $n$ kept in the expansion~\eqref{eq: reduced coordinates} and the third column denotes the dimensions of the PMM used for emulating the norm of the coefficients $\vec{a}$. The fourth and fifth columns denote the density locations $\rho$ where the EOS is evaluated to obtain the pressures which are the features $f_i(\alpha)$ in Eq.~\eqref{eq:H-matrix}, with square brackets representing equidistant points within the specified range in steps of 0.05 fm$^{-3}$. }
\label{tab:lambda-BB}
\startdata
Set 1 & 2 & 3 & (0.60, 0.75) & (0.45, 0.60)\\
Set 2 & 6 & 2 & [0.10-0.85] & [0.10-0.85]\\
Set 3 & 3 & 2 & [0.10-0.85] & [0.10-0.85]\\
\enddata
\end{deluxetable}

\begin{figure*}
    \centering
    \includegraphics[width=0.99\textwidth]{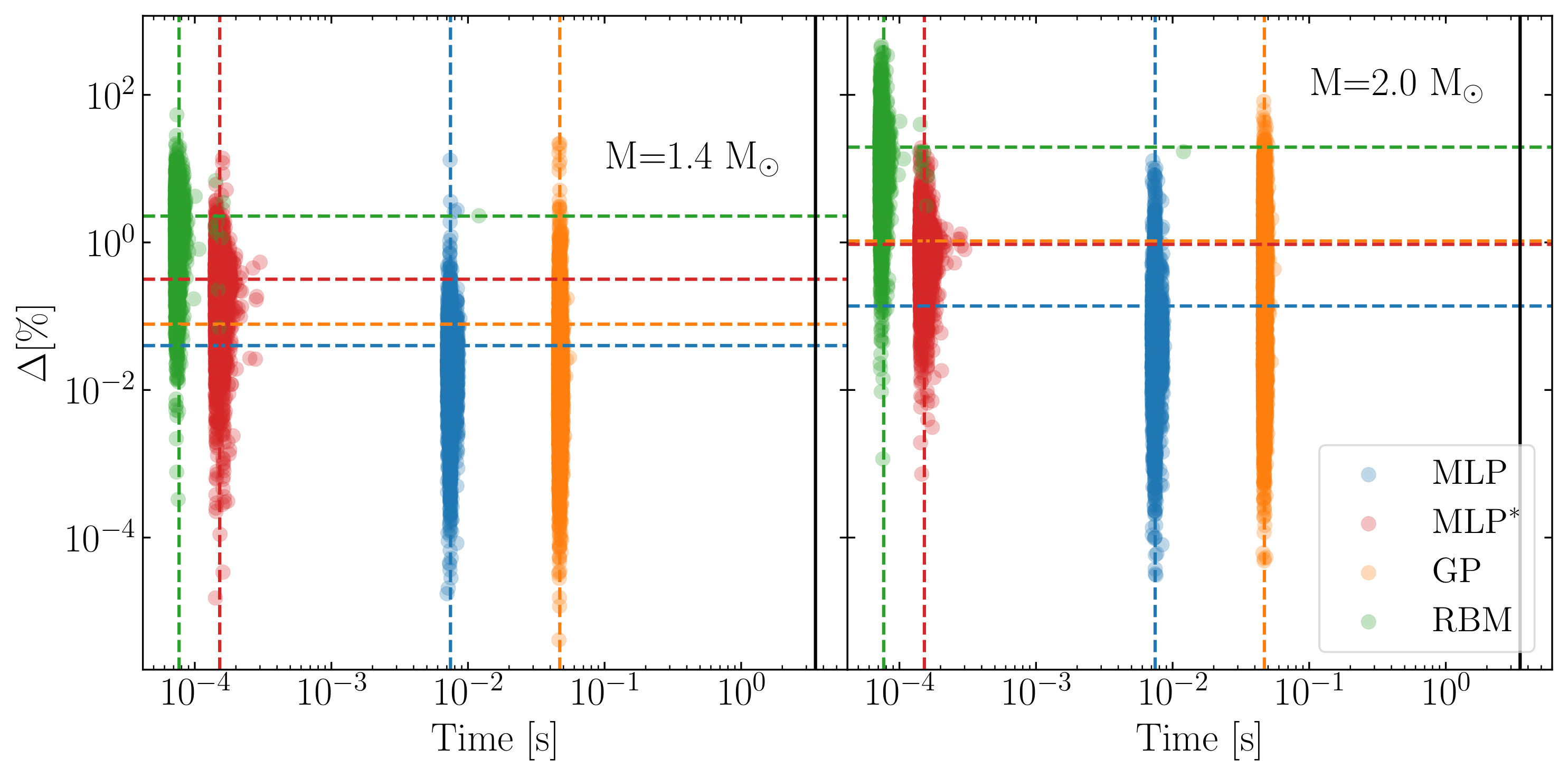}
    \caption{Same as Fig.~\ref{fig:CAT_s1}, but for the 5-parameter model.}
    \label{fig:CAT_s2}
\end{figure*}

To train the emulator and estimate the matrix elements of Eq.~\eqref{eq: eigensystem}, we use 2000 training samples for each EOS set. 
We then fit the matrix elements in \autoref{eq:H-matrix} for both the PCs and PMM emulators using the Levenberg-Marquardt least-squares routine~\citep{Levenberg,Marquardt},~performing ten fits on the same training samples but with different initial conditions. 
Specific to the PC emulator, we take the first normalized eigenvector corresponding to the largest eigenvalue of its H-matrix.
For the PMM, we emulate the normalization by taking the largest eigenvalue of its H-matrix.
We then validate each fit by calculating a $\chi^2$ of the resultant emulator with respect to the samples not used in the fit, and chose the emulator which minimizes this $\chi^2$.

In Fig.~\ref{fig:CV_RBM}, we show the validation performance of the RBM for the 5-parameter EOS model. 
The RBM can capture the general behavior of the full TOV solver, with the average of $\Delta_{1.4}$ being $2.4 \%$.
The largest outlier has a $\Delta_{1.4}\approx 60\%$ but only around 5 samples have such large uncertainties.
Note that at higher NS masses, the RBM sometimes predicts an unphysical, non-monotonous dependence of $\hat{\Lambda}$ on the NS mass since the RBM fails to emulate one of the PCs for the 5-parameter model.

\section{Comparison of the emulators}
\label{sec:CAT}

\begin{figure*}
    \centering
    \includegraphics[width=0.99\textwidth]{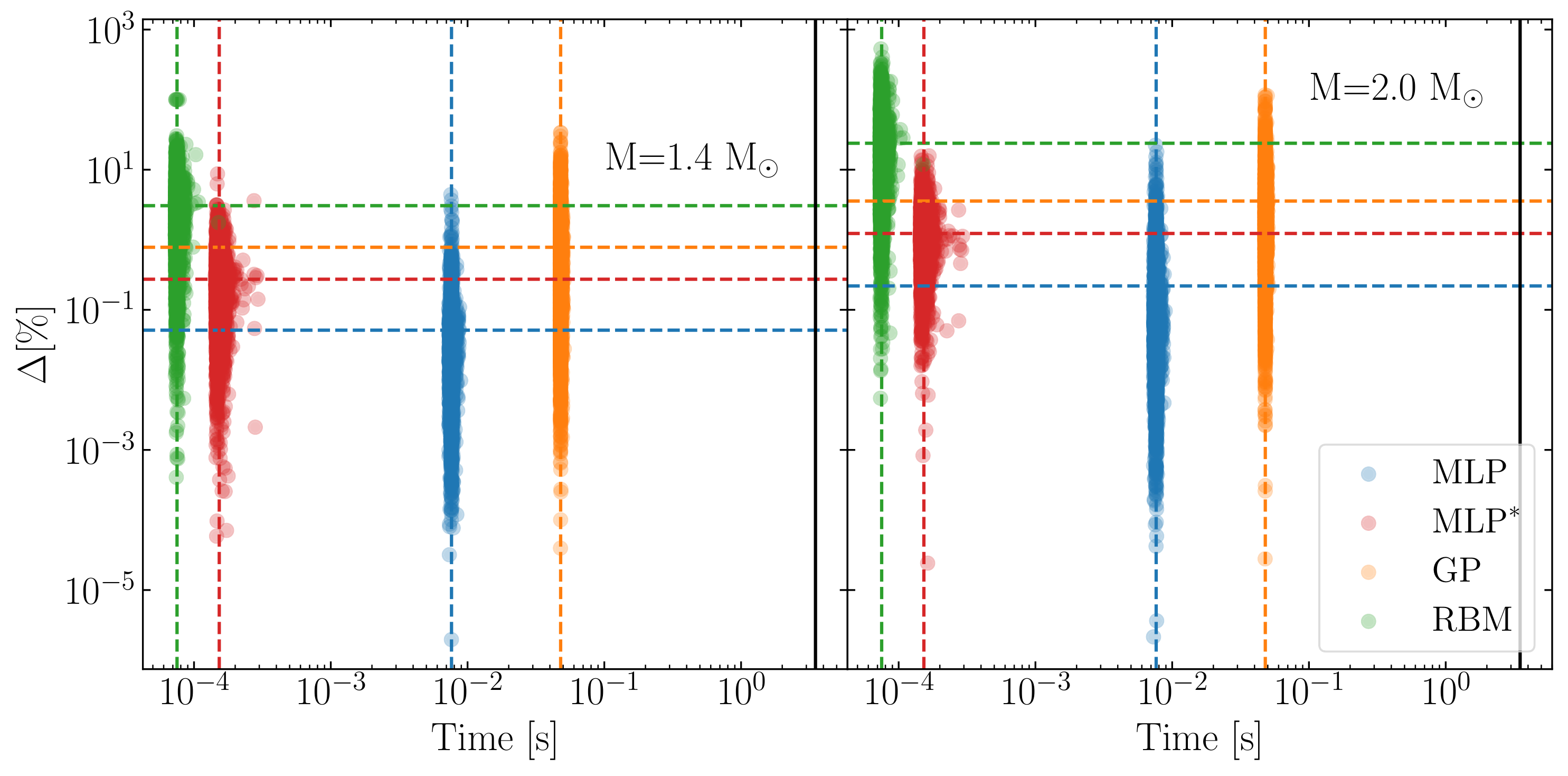}
    \caption{Same as Fig.~\ref{fig:CAT_s1}, but for the 10-parameter model.}
    \label{fig:CAT_s3}
\end{figure*}

In the following, we compare the performance of the three emulators in terms of both accuracy and computational speedup.
We show the computational accuracy vs. time plots in Figs.~\ref{fig:CAT_s1},~\ref{fig:CAT_s2}, and~\ref{fig:CAT_s3} for the 1-parameter, 5-parameter and 10-parameter models, respectively. 
The validation uncertainty is shown for stars with $1.4$M$_\odot$ ($2$M$_\odot$) in the left (right) panels. 
For a fair comparison, all emulators are evaluated on a single CPU core for each validation sample. 

From Figs.~\ref{fig:CAT_s1},~\ref{fig:CAT_s2}, and~\ref{fig:CAT_s3}, we observe that the MLP has the lowest average validation uncertainty among all emulators for all EOS models at both $1.4$M$_\odot$ and $2$M$_\odot$. 
This is followed by the MLP$^*$ and GP emulators which have comparable uncertainties, with MLP$^*$ generally showing lower average uncertainties, except for the left panel in Fig.~\ref{fig:CAT_s2}. 
In all cases, the RBM has the largest validation uncertainties.
Note that the comparison between MLP and MLP$^*$ allows us to gauge the impact of Bagging, see Sec.~\ref{sec:MLP}. 
Additionally, we examined the performance of the MLP emulator with respect to the number of members in the ensemble included in the Bagging algorithm. 
We found that the average validation uncertainty improves with the number of individual MLPs, before saturating at about 65 MLPs for the 5 and 10-parameter models. 
For the 1-parameter model, Bagging has a negligible impact, see Fig.~\ref{fig:CAT_s1}.  

The RBM offers the largest computational speedup compared to the full TOV solver, which takes $\approx 3.5$ seconds per sample.
The MLP$^*$ emulator is similarly fast and a factor of 100 faster than MLP, as expected. 
The GP emulator is the slowest among the emulators but still achieves over an order of magnitude speedup compared to the full solver. 
We conclude that for applications requiring high accuracy, the MLP emulator is the best choice. 
However, if computational speedup is the primary concern, the MLP$^*$ or RBM emulators may be more practical, albeit with increased emulator uncertainties. 
Given the typical computational resources used for BNS parameter inference runs~\citep{Biwer:2018osg,Ashton:2018jfp}, emulators which can output the $\Lambda$-M sequence within $\approx 10^{-2}$~s are sufficiently fast~\citep{Veitch:2014wba}. 
The required emulator accuracy is harder to estimate since this depends on the signal-to-noise ratio of the data and the desired precision on the EOS parameters. 
However, for typical events that are expected to be observed in the upcoming years, we expect that emulators with $\Delta  \approx 1\%$ would yield sufficiently accurate posterior distributions.    

As mentioned earlier, the primary factor limiting both speed and accuracy of the GP is its inability to handle large data sets. 
Although we found that Bagging improves the average uncertainty, the accuracy of the GP remains worse than both MLP and MLP$^*$. 
A possible solution might be the implementation of a sparse GP~\citep{Sparse_GP},   
but this is beyond the scope of this work. 
In contrast, the MLP emulator scales remarkably well with the size of the training set and we expect this improvement to continue for even larger training sets than those used in this work.
Increasing the size of the training set is, therefore, a simple and effective way of improving the MLP and MLP$^*$ emulators in the future. 

Finally, we note that the validation uncertainty is generally larger at $2$M$_\odot$ than $1.4$M$_\odot$. 
This is because the calculation of the percentage uncertainty requires dividing by the true tidal deformabilities which are much smaller ($\approx 10$) at $2$M$_\odot$ compared to their values at $1.4$M$_\odot$.

\section{Example analysis of a potential loud BNS observation}
\label{sec:loud_event}

To test our emulators in a  Bayesian parameter inference, we now study a fictitious loud BNS observation that could be detected by next-generation GW detectors, such as Cosmic Explorer~\citep{Evans:2021gyd}. 
We analyse the event using a simplified hierarchical approach~\citep{Landry:2022rxu} but will demonstrate our emulators in a full Bayesian parameter inference in future work.

The posterior distribution for a set of EOS parameters $\alpha$, given GW data $d$, can be written as 
\begin{align}
    P(\alpha | d) &\propto P( d |\alpha) P(\alpha) \nonumber \\
    &  \propto  P(\alpha) \bigg[ \int dm_1 dm_2 d\tilde{\Lambda} \ P( d |m_1,m_2,\tilde{\Lambda}) 
    \nonumber \\
    &\quad \quad \quad \quad \quad \quad \quad \quad P( m_1,m_2,\tilde{\Lambda} | \alpha ) \bigg]\,.
    \label{eq:bayes}
\end{align}
The function $P( m_1,m_2,\tilde{\Lambda} | \alpha)$ is determined by a given EOS represented by $\alpha$. 
The variables $m_1$ and $m_2$ are the component masses and $\tilde{\Lambda}$ is the combined dimensionless tidal deformability of the binary~\citep{Hinderer:2009ca}.
Here, we fix the component masses of our injected event to $1.4$M$_\odot$. 
Then, the marginal likelihood of the BNS parameters $P( d |m_1,m_2,\tilde{\Lambda}) \equiv \mathcal{L}(m_1,m_2,\tilde{\Lambda})$  has the form
\begin{eqnarray}
    \mathcal{L}(m_1,m_2,\tilde{\Lambda}) = \delta(m_1-1.4) \delta(m_2-1.4) \mathcal{L}(\tilde{\Lambda})\,.
\end{eqnarray}
For the factorized likelihood $\mathcal{L}(\tilde{\Lambda})$, we assume a simple Gaussian distribution centered at $\tilde{\Lambda}=500$ with a standard deviation of $25$ which corresponds to a $5\%$ uncertainty. 
We have checked that the qualitative conclusions of this section do not change upon varying the mean and standard deviation of the Gaussian distribution.  
We then evaluate the posterior distribution on $\alpha$ using~\autoref{eq:bayes}. 
In practice, this is done by drawing a large number of samples from the prior $P(\alpha)$ and assigning each a likelihood weight proportional to the integral within the square brackets in~\autoref{eq:bayes}. 
Note that the integral simplifies drastically under the above approximations. 

\begin{figure*}
    \centering
    \includegraphics[width=0.99\columnwidth]{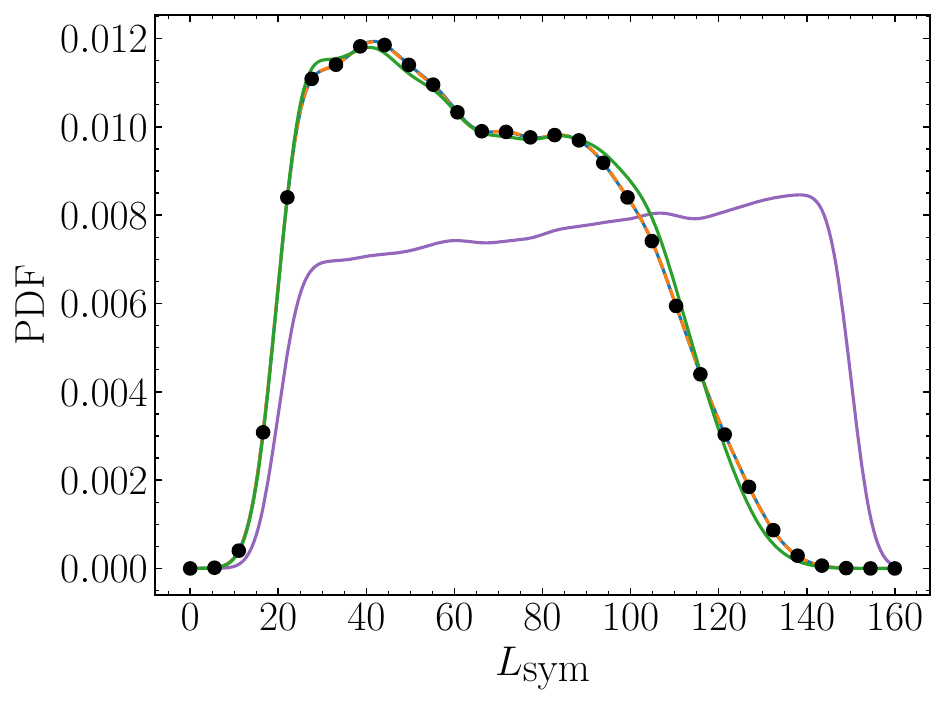}
    \includegraphics[width=0.99\columnwidth]{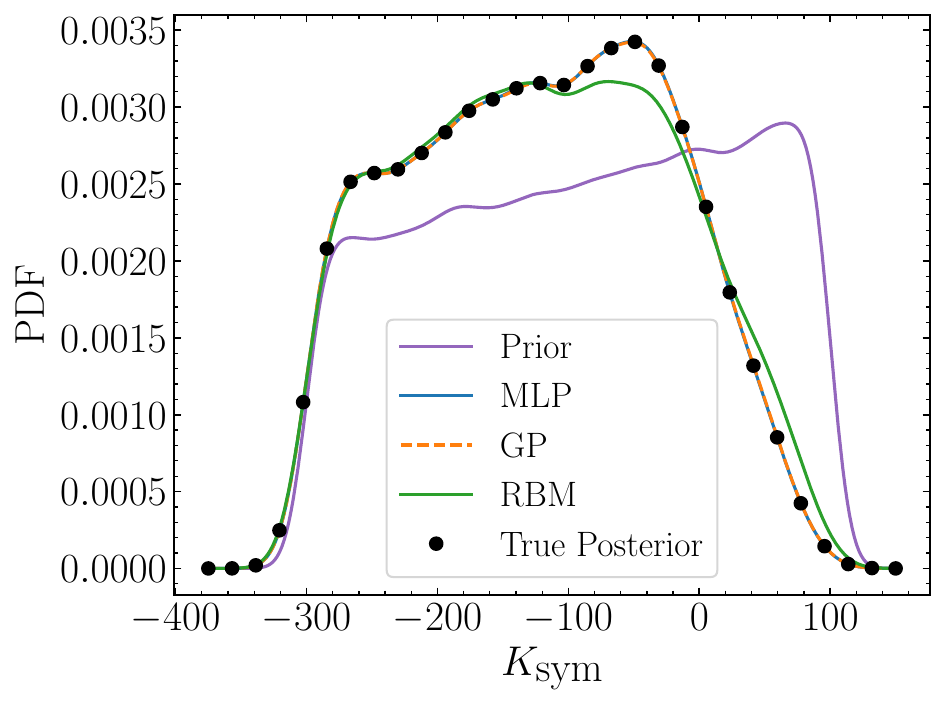}
    \caption{PDFs on two EOS model parameters, computed using the 5-parameter model, for a fictitious high-precision measurement of the tidal deformability of a $1.4$M$_\odot$ NS described in the text.
    The PDFs for all emulators matches well with that obtained using the full TOV solver (black circles).}
    \label{fig:L_sym}
\end{figure*}

For the samples drawn from the prior $P(\alpha)$, we employ our set of 100K validation samples.
Each validation sample corresponds to the true $\Lambda-$~M curve, calculated with the full TOV solver, as well as the predicted $\hat{\Lambda}-$~M curve calculated using the emulators. 
Therefore, we obtain the posterior on $\alpha$ calculated from both the full solver as well as the emulators. 
Our results for the 5-parameter model are shown in Fig.~\ref{fig:L_sym} where we show the marginalized 1-d posterior distribution on two NEPs, $L_\text{sym}$ and $K_\text{sym}$. 
We find that the emulated PDFs for the GP and MLP match perfectly with the PDF of the full TOV solver. 
We have checked that this is the case for all $5$ parameters of the EOS model.
Results for the 1-parameter and 10-parameter EOS models are qualitatively similar.
While the RBM does not exactly reproduce the true PDF, the agreement is nevertheless very good. 
Overall, our results demonstrate the potential of our emulators for enhancing Bayesian parameter inference of BNS observations.    

\section{Conclusion}
\label{sec:conclusion}

To accelerate the inference of nuclear physics parameters from astrophysical NS observations, we have developed three different algorithms for emulating solutions to the TOV equations.
We have focused on emulating the tidal deformability, as this is the key quantity for GW data analyses of BNS mergers.
However, our framework can easily be extended to emulate other astrophysical NS observables such as the NS radii, moments of inertia, baryonic masses, Kepler masses, etc. 
We have developed our emulators for three different EOS models, two of which are capable of incorporating constraints from nuclear theory and experiment while simultaneously being agnostic enough to allow for phase transitions at high densities. 
Our MLP emulator proves to be the most accurate, while the RBM offers the best speedup. 
All emulators are capable of predicting the full $\Lambda - $~M sequence within $0.1$ seconds.
Finally, we have shown that the PDFs on EOS model parameters predicted by our emulators match very well with the true solution for a loud mock BNS signal with a 5\% measurement uncertainty. 
Given these encouraging results, the work presented in this paper and further developments to our emulation strategies will pave the way for future studies aiming to extract the nuclear EOS from observations of BNSs and other astrophysical sources including NSs.

\section{acknowledgements}
We thank Edgard Bonilla, Diogenes Figueroa, and Kyle Godbey for useful conversations and insights on the use of data-driven RBMs, and Patrick Cook and Danny Jammooa for useful discussions about PMM implementations.
B.T.R. was supported by the Laboratory Directed Research and Development program of Los Alamos National Laboratory under project number 20230785PRD1.
R.S. and D.A.B. acknowledge support from the Nuclear Physics from Multi-Messenger Mergers (NP3M) Focused Research Hub which is funded by the National Science Foundation under Grant Number 21-16686.
R.S. acknowledges support from the Laboratory Directed Research and Development program of Los Alamos National Laboratory under project number 20220541ECR.
S.D., C.L.A., and I.T. were supported by the Laboratory Directed Research and Development program of Los Alamos National Laboratory under project number 20230315ER. 
I.T. was also supported by the U.S. Department of Energy, Office of Science, Office of Nuclear Physics, under contract No.~DE-AC52-06NA25396, and by the U.S. Department of Energy, Office of Science, Office of Advanced Scientific Computing Research, Scientific Discovery through Advanced Computing (SciDAC) NUCLEI program. 
P.G. was supported by the National Science Foundation CSSI program under Grant No. OAC-2004601 (BAND Collaboration~\citep{phillips2021get}), and Michigan State University and the Facility for Rare Isotope Beams. 
C.C. acknowledges support from NSF award PHY-2309356.

\bibliography{bib}

\end{document}